\documentstyle[epsfig]{mn} 
\begin{document}

\title
[Lensing and the SZ effect]
{Gravitational lensing and the Sunyaev--Zel'dovich effect in the 
millimetre/submillimetre waveband} 
\author
[A. W. Blain]
{
A. W. Blain\\
Cavendish Laboratory, Madingley Road, 
Cambridge, CB3  0HE.
}
\maketitle 

\begin{abstract}
The intensity of the cosmic microwave background radiation in the fields of 
clusters of galaxies is modified by inverse Compton scattering in the hot 
intracluster gas -- the Sunyaev--Zel'dovich (SZ) effect. The effect is 
expected to be most pronounced at a frequency of about 350\,GHz (a wavelength 
of about 800\,$\mu$m), and has been detected in the centimetre and millimetre 
wavebands. In the millimetre/submillimetre waveband, the gravitationally-lensed 
images of distant dusty star-forming galaxies in the background of the cluster 
are predicted to dominate the appearance of clusters on scales of
several arcseconds, and could confuse observations of the SZ effect at 
frequencies greater than about 200\,GHz (wavelengths shorter than about 
1.5\,mm). Recent observations by Smail, Ivison \& Blain (1997) confirm that a 
significant population of confusing sources are present in this waveband.
Previous estimates of source confusion in observations of the 
millimetre/submillimetre-wave SZ effect did not include the effects of lensing by 
the cluster, and so the accuracy of such measurements could be lower
than expected. Source subtraction may be required in order to measure the SZ
effect accurately, and a careful analysis of the results of an ensemble of SZ 
measurements could be used to impose limits to the form of evolution of distant 
dusty star-forming galaxies.
\end{abstract}


\section{Introduction} 

Gravitational lensing can have a more significant effect on the population of 
galaxies detected in the millimetre/submillimetre waveband as compared with 
other wavebands; Blain (1996a,b; 1997a,b -- Paper\,1; 1997c, 1998) gives details 
of the effects of lensing by both galaxies and clusters. Here we consider the 
effects of the lensed images of background galaxies in 
millimetre/submillimetre-wave
observations of clusters on arcminute angular scales that are suitable to detect 
the modifications to the intensity of the cosmic microwave background radiation 
(CMBR) due to the Sunyaev--Zel'dovich (SZ) effect (Sunyaev \& Zel'dovich 1980; 
Rephaeli 1995b). Lensed images are not expected to be resolved on these scales, 
but they could produce significant source confusion because of their uncertain 
spatial distribution in the observing beam. Accurate estimates of this confusion 
could be necessary in order to interpret the results of observations of the SZ 
effect. The importance of source confusion in this waveband has recently been 
confirmed by the 850-$\mu$m observations of lensed images in the clusters
ABell\,370 and Cl2244$-$02 (Smail, Ivison \& Blain 1997 -- SIB). The effects of 
millimetre/submillimetre-wave source confusion in the absence of gravitational 
lensing are discussed by Blain, Ivison \& Smail (submitted -- BIS). 

The principal sources of millimetre/submillimetre-wave radiation from clusters 
are the SZ effect and the lensed images of distant dusty background galaxies. 
Other sources are star-forming or active galaxies within the cluster, the 
Rees--Sciama (RS) effect (Rees \& Sciama 1968; Quilis, Ib\'a\~nez \& Saez 1995) and 
primordial CMBR anisotropies. We discuss the properties and relative importance of 
all five sources in Section~2, and estimate the source confusion noise expected 
due to the lensed images of distant galaxies in Section~3. In Section~4 the 
consequences of this confusion for 
millimetre/submillimetre-wave observations of the SZ effect are discussed, and 
the prospects for using the same observations to detect the signature of galaxy 
evolution in the population of lensed images are assessed. Throughout this 
paper we assume that the density parameter $\Omega_0=1$ and that Hubble's 
constant $H_0 = 50$\,km\,s$^{-1}$\,Mpc$^{-1}$. 

\section{Sources of millimetre/submilli- metre-wave radiation in clusters} 

\subsection{The Sunyaev--Zel'dovich effect} 

The SZ effect describes the modifications to the spectrum of the CMBR 
due to inverse Compton scattering in the hot X-ray emitting gas bound to 
clusters of galaxies. It is usually separated into two components: a thermal effect,
which describes the net energy gain of CMBR photons scattered by hot gas 
moving with the Hubble flow; and a kinematic effect, which describes the net 
Doppler shift of the photons scattered in a cluster with a non-zero peculiar 
velocity.

If the scattering gas is non-relativistic, then the thermal SZ effect produces a 
characteristic increment and decrement to the CMBR intensity above and below 
a frequency of 217\,GHz respectively. The magnitude of this effect is independent 
of the redshift of the scattering gas and is determined by a single dimensionless 
parameter $y=\tau \sigma_{\rm T} k_{\rm B} T_{\rm e} / ( m_{\rm e} c^2 )$, in which 
$\tau$ is the scattering optical depth of the cluster and $\sigma_{\rm T}$, 
$T_{\rm e}$ and $m_{\rm e}$ are the electron scattering cross section, 
temperature and mass respectively. The $y$-parameter is typically several times 
$10^{-4}$ in a rich cluster of galaxies (Rephaeli 1995b), and 
$\vert y \vert < 1.5 \times 10^{-5}$ over the whole sky (Fixsen et~al. 1996). The 
magnitude of the 
kinematic SZ effect, which is expected to be smaller than the thermal SZ effect in 
a typical cluster, is proportional to $\tau v_{\rm r}$, where $v_{\rm r}$ is the 
peculiar velocity of the scattering gas. If the scattering gas is relativistic, then 
the form of the thermal effect is no longer described completely by the 
$y$-parameter, and the values of both $\tau$ and $T_{\rm e}$ affect the 
spectrum of the CMBR independently (Rephaeli 1995a).

The spectrum of both the thermal and kinematic SZ effects expected for two 
clusters, both with $\tau = 10^{-2}$ and $v_{\rm r} = 1000$\,km\,s$^{-1}$, are 
shown in Fig.\,1. One cluster contains non-relativistic electrons with 
$k_{\rm B} T_{\rm e} < 5$\,keV and has $y=1.4\times10^{-4}$; the other 
contains relativistic electrons with $k_{\rm B} T_{\rm e} = 10$\,keV. Limits to the 
intensity of the diffuse extragalactic background radiation from the FIRAS 
instrument on the {\it COBE} satellite (Puget et al. 1996; Fixsen et al. 1996) are 
also shown.

The centimetre-wave SZ effect has been detected by Birkinshaw (1990), 
Uyaniker et al. (1997) and Myers et al. (1997) at frequencies of 20.3, 10.55 and 
32\,GHz respectively using single-antenna telescopes, and by Jones et al. (1993) 
and Carlstrom, Joy \& Grego (1996) using interferometers at 15 and 30\,GHz 
respectively. Recently, Grainge et~al. (1996) produced the first image of the 
SZ effect at 15\,GHz. In the millimetre waveband, the SZ decrement has been 
detected near the peak of the CMBR spectrum at 2.2\,mm (Wilbanks et~al. 1994),
using the SuZIE bolometer array receiver at the Caltech Submillimeter
Observatory (Holzapfel et al. 1997), and Andreani et~al. (1996) have detected 
the SZ increment at 1.2\,mm. Silverberg et al. (1997) recently detected the  
submillimetre-wave SZ effect in the direction of the Coma cluster at four 
frequencies but low significance using the MSAM balloon-borne experiment. 

Current observations of the SZ effect are consistent 
with a cored isothermal surface brightness distribution, which depends on 
angular radius $\theta$ as $[1 + (\theta / \theta_{\rm c})^2 ]^{-1/2}$. The rich 
clusters Abell 2218 and Abell 2163 at redshifts $z_{\rm c} = 0.171$ and 0.201 have 
core radii $\theta_{\rm c} \simeq 50$ and 70\,arcsec respectively (Squires et al. 
1996; Elbaz, Arnaud \& B\"ohringer 1995).

\begin{figure}
\begin{center}
\epsfig{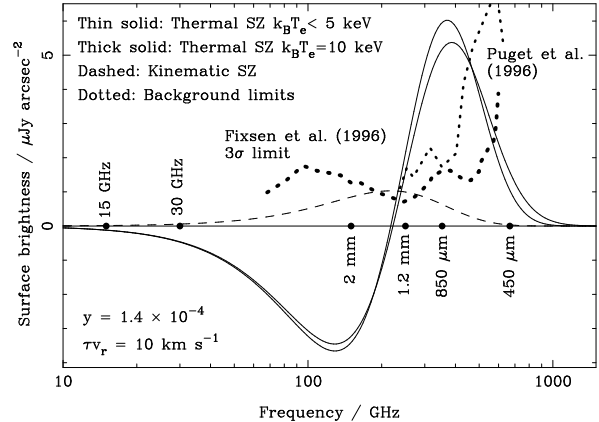}
\end{center}
\caption{Predictions of the spectrum of the thermal and kinematic SZ effects in a
typical rich cluster. The frequency of the peak in the CMBR spectrum is 160\,GHz. 
Observed limits to the intensity of diffuse extragalactic background radiation and 
the frequencies of important observing bands are also shown. 
}
\end{figure} 

\subsection{Lensed images of background galaxies} 

The properties of the lensed images of distant dusty star-forming galaxies
lensed by a rich cluster at a redshift $z_{\rm c} = 0.17$ were discussed in 
Paper~1, and are expected to have flux densities in the range 1 to 10\,mJy at 
850\,$\mu$m, comparable to or larger than the flux densities of any star-forming 
galaxies within the lensing cluster. Distant galaxies are expected to be relatively 
bright in the millimetre/submillimetre waveband because the {\it K}-corrections 
due to redshifting the dust emission spectra of distant galaxies are expected to 
be large and negative as compared with other wavebands (Blain \& Longair 
1993a). Hence, faint galaxies detected in the millimetre/submillimetre waveband 
are typically expected to have redshifts much larger than $z_{\rm c}$, and so, 
even if $z_{\rm c}$ were increased substantially, a large surface density of 
background galaxies and lensed images would still be expected.

SIB have recently detected examples of such distant dusty galaxies in the fields 
of the clusters A370 and Cl2244$-$02 at $z_{\rm c}=0.37$ and 0.33 respectively. 
The surface density of these objects is larger than that predicted previously 
(Blain \& Longair 1996), but is consistent with the observed background radiation 
intensity in the far-infrared waveband (Puget et al. 1996) and with recent 
observations of the Hubble Deep Field at a wavelength of 2.8\,mm 
(Wilner \& Wright 1997). Even in the absence of 
gravitational lensing these sources are expected to contribute a significant 
amount of confusion noise to some observations in the millimetre/submillimetre 
waveband (BIS). 

\subsection{Other sources of radiation} 

The far-infrared radiation emitted by dust in star-forming galaxies within a 
cluster could produce a detectable flux density in the submillimetre waveband. 
The fraction of blue star-forming galaxies in clusters tends to increase 
with increasing redshift, described as the Butcher--Oemler (BO) effect 
(Butcher 1978, Lavery \& Henry 1994, Couch et~al. 1994), and would correspond 
to an increase in the luminosity of these galaxies in the far-infrared waveband.

The Rees--Sciama (RS) effect is expected to produce a signal with a 
spectrum similar to that of the CMBR. If the density of the core of a cluster 
evolves during its light crossing time, then the gravitational blueshift imposed on 
a CMBR photon as it falls into the cluster is not matched by the gravitational 
redshift imposed as it climbs out, and so the spectrum of the CMBR is modified. 
Quilis, Ib\'a\~nez \& Saez (1995) have calculated the largest possible signal due 
to this effect. 

Gravitational lensing of primordial anisotropies in the CMBR would lead to a 
spatially-correlated millimetre/submillimetre-wave background signal on scales 
similar to the Einstein radius of the lensing cluster (Paper\,1).

\subsection{The relative importance of these sources}

The frequency-dependent surface brightness of the SZ increment to the CMBR 
intensity in the core of a cluster with $y=1.4\times10^{-4}$ should peak at about 
6\,$\mu$Jy\,arcsec$^{-2}$ at a wavelength of about 850\,$\mu$m (Fig.\,1). In
comparison, the largest values of surface brightness of resolved lensed images 
and galaxies in the cluster at $z_{\rm c}=0.171$ that was modeled in Paper\,1 
were about 60 and 6\,$\mu$Jy\,arcsec$^{-2}$ respectively. The RS effect and 
primordial CMBR anisotropies are expected to produce 
much smaller surface brightnesses of order 300 and 30\,nJy\,arcsec$^{-2}$ 
respectively at their maximum intensities (Paper\,1). 
 
The millimetre/submillimetre-wave flux densities of any dusty star-forming galaxies 
within a cluster are expected to be considerably smaller than those of the lensed 
images of distant galaxies unless the galaxies are in a cluster at a 
low redshift, that is if $z_{\rm c}$ is significantly less than 0.1 (Paper\,1). The 
spectral energy distributions of distant lensed images should be flatter than 
those of cluster galaxies, and so at higher frequencies the cluster galaxies
should be relatively brighter as compared with the lensed images;  
however, the lensed images are still expected to be brighter than the cluster 
galaxies throughout the millimetre/submillimetre waveband. The typical flux
densities of the lensed images should approximately follow the envelope to the 
observed limits to the background radiation intensity shown in Fig.~2. 

Lensed images are expected to dominate the appearance of clusters on 
arcsecond scales; however, on larger angular scales the surface brightness 
distribution of the SZ effect becomes more significant, and the SZ signal is 
expected to dominate in observations on arcminute scales at 850\,$\mu$m.

\section{Estimating source confusion} 

\subsection{Introduction} 

Source confusion is the uncertainty introduced into the results of an observation 
by the flux densities of unresolved sources that lie at unknown positions in the 
observing beam. Source confusion due to extrapolated and inferred populations of 
distant dusty galaxies has been estimated by Franceschini et al. (1989, 1991), 
Helou \& Beichman (1990), Toffolatti et al. (1995) and Gawiser \& Smoot (1997). 
BIS have recently made the first estimates to be based on the results of direct 
submillimetre-wave observations. Fischer \& Lange (1993) discussed the effects of 
unlensed source confusion on observations of the SZ effect, and estimated that 
the confusion noise at a wavelength of 850\,$\mu$m in a 1-arcmin observing 
beam should be about 20 times smaller than the signal of the SZ effect from a 
cluster with $y=10^{-4}$. BIS estimate that source confusion at 850\,$\mu$m is 
actually about a factor of 7 times more severe than this earlier value.

Confusion noise in the field of a cluster should be increased by gravitational 
lensing, because both the mean separation and flux density of bright lensed 
images are expected to be larger than those of unlensed background galaxies. In 
the following sections we estimate the size of this increase and discuss its 
effects on observations of the SZ effect. In order to make an accurate 
measurement of the SZ effect, the flux densities of confusing sources may need
to be subtracted from the observed signal. This is usually achieved by observing 
the cluster to the same limiting flux density but at a finer angular resolution, in 
order to detect the confusing galaxies while resolving out the SZ signal. 
Loeb \& Refregier (1997) have shown that this 
procedure could introduce a subtle bias into measurements of the SZ effect. 
Because lensed images are typically brighter than unlensed galaxies, a larger 
fraction of their total number can be detected at any flux density limit as 
compared with unlensed galaxies. The images are expected to be more 
strongly magnified in the inner few arcminutes of the field of a cluster, and so 
source subtraction is expected to be more efficient there as compared with the 
outer regions of the field. It is this difference in efficiencies that introduces 
the systematic error into the SZ measurement. However, this effect is 
not of immediate concern in the millimetre/submillimetre waveband, because 
accurate source subtraction will be impossible in this waveband until large 
interferometer arrays are available (Brown 1996; Downes 1994, 1996). The current 
accuracy of millimetre-wave observations of the SZ effect is limited by the 
uncertain flux densities of all the confusing galaxies and not just by their 
incomplete subtraction.  

\subsection{The procedure} 

The effects of gravitational lensing on source confusion was estimated by 
combining several different models of galaxy evolution (Blain \& Longair 1996) 
with a model of lensing by clusters (Paper 1). A large set of random distributions 
of background galaxies and their lensed images were derived, and their surface
brightness distributions were convolved with Gaussian beams on various scales 
in order to produce distributions of the flux densities due to both lensed and 
unlensed sources in the beams. The widths of these distributions were then 
used to predict the source confusion noise expected with and without
gravitational lensing.

\subsubsection{Modeling galaxy formation and evolution} 

Three models of the evolution of the population of dusty star-forming galaxies 
are included in this paper; models I1 and I2 are based on pure luminosity 
evolution of the {\it IRAS} luminosity function (Saunders et al. 1990), and model H 
is based on a model of mergers in an hierarchical clustering scheme (Blain \& 
Longair 1993a,b) that is derived from the Press--Schechter formalism (Press \& 
Schechter 1974). The models are described in more detail by Blain \& Longair 
(1996), in which model I1 was labelled model 3. In model I1 pure luminosity
evolution is by a factor $(1+z)^3$ when $z\le2$ and by a factor 27 when $2<z\le5$.
In model I2, the $(1+z)^3$ form of evolution extends to a larger redshift
$z=2.6$, and the evolution factor is 46.7 when $2.6<z\le7$. Model I2 was used to 
describe the observed counts of SIB. The background radiation intensities predicted 
by each model are shown in Fig.\,2 and are consistent with the observed limits to 
the diffuse background radiation intensity (Puget et al. 1996; Fixsen et al. 1996). 

\subsubsection{Modeling the effects of lensing} 

A ray-tracing method for determining the distribution and properties of the 
lensed images of galaxies behind a cluster was demonstrated in Paper~1. 
A spherical model of a lensing cluster with properties similar to those of 
Abell~2218 was used (Natarajan \& Kneib 1996); that is, with a velocity dispersion 
$\sigma_{\rm V}=1360$\,km\,s$^{-1}$ at a redshift $z_{\rm c} = 0.171$ 
(Kneib et al. 1996). The effects that different values of $\sigma_{\rm V}$ and 
$z_{\rm c}$ have on the predicted source confusion are discussed in 
Section\,3.4. 

\begin{figure}
\begin{center}
\epsfig{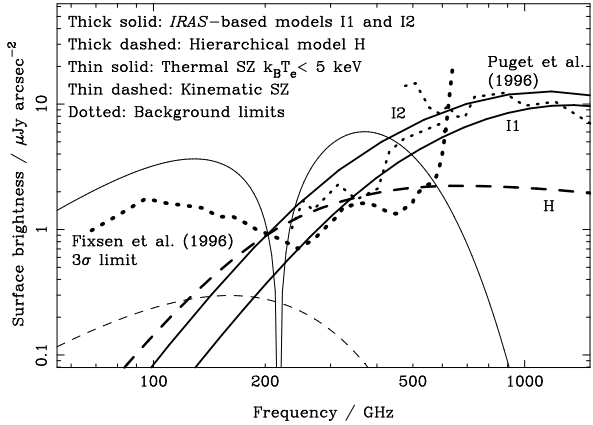}
\end{center}
\caption{
The predicted diffuse background intensities in three models of galaxy formation 
(Table\,1), limits to the intensity of diffuse background emission and the spectrum 
of the SZ effect (Fig.\,1). Note that the diffuse background intensities are much 
larger than the expected level of source confusion noise.  
}
\end{figure}
 
\subsubsection{Combining the models}

The source confusion with and without gravitational lensing was estimated from 
a large number of surface brightness distributions at chosen frequencies in the 
millimetre/submillimetre waveband, which were set up in 10-arcmin-wide fields 
using galaxies drawn at random from the luminosity functions describing 
the galaxy distribution expected in models I1, I2 and H above. The corresponding 
surface brightness distribution of the lensed images of these galaxies in each
field were then determined using the lensing model. Each lensed and 
unlensed surface brightness distribution was then convolved with a set of 
Gaussian beams, with full-width-half-maximum (FWHM) beam-widths 
$\theta_{\rm b}$ between 5 and 300\,arcsec in order to produce values of the flux 
density of unresolved sources on each angular scale. After repeating this process 
many times the width of the resulting distributions of flux density values, 
$\sigma_{\rm I}(\theta_{\rm b})$ and $\sigma_{\rm S}(\theta_{\rm b})$ for lensed 
and unlensed sources respectively, were determined and used to estimate the 
confusion noise. These widths were defined as half the range of flux density that 
included 60 per cent of the simulated flux densities. Any difference in the mean 
flux density in the simulated lensed and unlensed fields gives an estimate of the 
residual signal expected due to lensing; this difference was always found to be 
smaller than $\sigma_{\rm I}$ and $\sigma_{\rm S}$.

\begin{figure}
\begin{center}
\epsfig{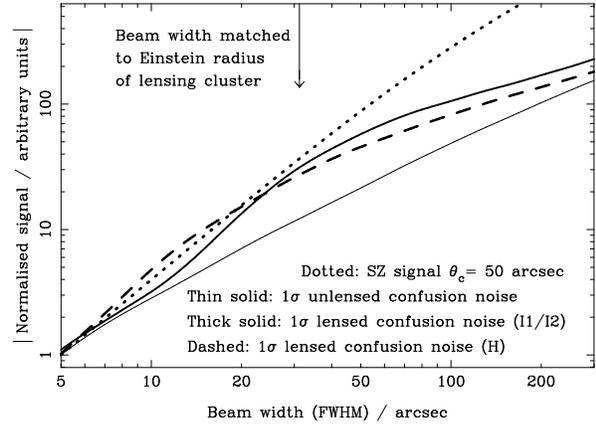}
\end{center}
\caption{The relative amplitude of the 1$\sigma$ confusion noise expected with 
and without lensing in a beam-switched observation of the SZ effect due to 
distant dusty star-forming galaxies as a function of beam-width. The predicted
amplitudes are different in model H as compared with models I1 and I2. An 
arbitrarily-normalised SZ signal is also shown.
}
\end{figure} 

The SZ effect can be detected by either making a fully-sampled image of the field
of a cluster or in a `beam-switched' observation that measures the differential 
signal between a beam that is centred on the core of the cluster and a reference 
beam that is offset by several beam-widths. The confusion noise expected in a 
beam-switched observation is given by the quadrature sum of the values of 
confusion noise in each beam, which are both given by $\sigma_{\rm S}$ if
the effects of lensing are neglected, and by $\sigma_{\rm I}$ and $\sigma_{\rm S}$ 
if lensing is included. In an imaging observation, estimates of confusion noise with
and without lensing are given directly by $\sigma_{\rm I}$ and $\sigma_{\rm S}$ 
respectively.

\subsection{The results} 

The estimates of confusion noise presented here were derived from simulations in 
240 random fields at wavelengths of 2\,mm, 1.2\,mm, 850\,$\mu$m and 
450\,$\mu$m. The dependence of confusion noise on beam-width was found to 
be very similar at each wavelength; Fig.\,3 shows the combined profile of the 
results at all four wavelengths, with and without lensing for model H and for 
models I1 and I2. A typical predicted profile of the SZ signal is also shown. 
The increase in source confusion due to gravitational lensing is most pronounced 
when the beam-width $\theta_{\rm b}$ is matched to the Einstein radius of the 
lensing cluster $\theta_{\rm E}$, that is when 
$\theta_{\rm b} \sim 2\theta_{\rm E}$. In these calculations 
$\theta_{\rm E} \simeq 16$\,arcsec (Paper~1), and so source confusion is 
predicted to increase most significantly, by a factor of about 3, at 
$\theta_{\rm b} \simeq 32$\,arcsec. The predicted increase in confusion is smaller 
in larger beams, and tends gradually to its unlensed value when 
$\theta_{\rm b} \gg \theta_{\rm E}$.

\begin{table*}
\begin{minipage}{175mm}
\caption{Estimates of the 1$\sigma$ confusion noise expected for SZ 
measurements due to lensed and unlensed background galaxies. The
$y$-parameter and core radius of the cluster are assumed to be 
$1.4\times10^{-4}$ and 50\,arcsec respectively. The estimates are uncertain to 
within about 10 per cent.}
\begin{tabular*}{170mm}{p{18mm} p{13mm} p{25mm} p{55mm} p{40mm}}
Frequency /& FWHM & SZ & 1-$\sigma$ confusion noise: & 
Ratio of SZ signal to confusion\\
GHz (Wave-& beam & signal & lensed (unlensed) / mJy & 
noise: lensed (unlensed) \\
\noalign{\vskip -1pt}
\end{tabular*}
\begin{tabular*}{170mm}{p{18mm} p{13mm} p{13mm} p{16mm} p{16mm} p{21mm} 
p{15mm} p{15mm} p{16mm}}
length / $\mu$m)&/ arcsec &/ mJy & ModeI I1 & Model I2 & Model H & ModeI I1 & 
Model I2 & Model H\\
\noalign{\vskip 4pt}
\end{tabular*}
\begin{tabular*}{170mm}{p{18mm} p{13mm} p{13mm} p{2mm} p{10mm} p{2mm}
p{9.5mm} p{2mm} p{15mm} p{2mm} p{9mm} p{2mm} p{8.5mm} p{2mm} p{10mm} }
150 (2000)& 60& -11.7 & 
0.54 & (0.24) & 0.77 & (0.46) & 0.46 & (0.17) & 22& (49) & 15 & (25) & 25 & (69) \\
 & 120& -36.3 & 
0.93 & (0.54) & 1.48 & (0.90) & 0.78 & (0.36) & 39 & (68) & 25 & (40) & 47 & (100)\\
\noalign{\vskip 3pt}
250 (1200) & 60& 7.32 & 
3.24 & (1.42) & 4.51 & (2.57) & 1.42 & (0.55) & 2.3 & (5.2) & 1.6 & (2.9) & 5.2 & (13)\\
 & 120& 22.7 & 
5.58 & (3.13) & 8.79 & (5.23) & 2.42 & (1.14) & 4.1 & (7.3) & 2.6 & (4.3) & 9.4 & (20)\\
\noalign{\vskip 3pt}
350 (850) & 60& 20.5 & 
9.39 & (3.89) & 12.4 & (6.88) & 2.26 & (0.93) & 2.2 & (5.3) & 1.7 & (3.0) & 9.1 & (22)\\
 & 120& 63.6 & 
16.0 & (8.56) & 24.6 & (14.3) & 3.94 & (1.98) & 4.0 & (7.4) & 2.6 & (4.4) & 16 & (32)\\
\noalign{\vskip 3pt}
670 (450) & 60& 3.61 & 
41.3 & (14.3) & 44.0 & (23.2) & 4.22 & (2.46) & 0.09 & (0.25) & 0.08 & (0.16) & 0.86 & 
(1.5)\\
 & 120& 11.2 & 
66.6 & (31.9) & 90.5 & (49.1) & 7.96 & (5.76) & 0.17 & (0.35) & 0.12 & (0.23) & 1.4 & 
(1.9)\\
\end{tabular*}
\end{minipage}
\end{table*}  

The modifications to the confusion noise due to lensing take this form because 
the most strongly magnified images are formed at angular radii close to 
$\theta_{\rm E}$; most images at smaller radii are demagnified, and images at 
larger radii are magnified by progressively smaller amounts. In a beam smaller 
than the Einstein radius, lensing modifies the confusion noise slightly, but as 
$\theta_{\rm b}$ increases towards $2 \theta_{\rm E}$, a steadily increasing 
fraction of the flux density from strongly magnified images at $\theta_{\rm E}$ is 
included in the observing beam, and so the confusion noise increases. As 
$\theta_{\rm b}$ increases above $\theta_{\rm E}$, the less strongly magnified 
images at radii larger than $\theta_{\rm E}$ contribute a steadily increasing 
fraction of the detected flux density; therefore, the contribution of the
strongly magnified images at radii near $\theta_{\rm E}$ is steadily 
diluted, and so the confusion noise tends gradually to its unlensed value.

The confusion noise expected in each observing band and model of galaxy 
evolution is shown in Table\,1 and Fig.\,4. The results for beam-widths 
of 1 and 2\,arcmin are presented in Table\,1, while in Fig.\,4 the results in a 
1-arcmin beam are compared with earlier estimates. The estimates of unlensed
confusion noise are significantly larger than those of Helou \& Beichman (1990) 
and Fischer \& Lange (1993), but are more similar to those of Franceschini et al. 
(1991). They are in good agreement with observationally-based estimates (BIS). 
Confusion noise is expected to dominate the SZ signal at a wavelength of 
450\,$\mu$m and to be very significant at wavelengths of 850\,$\mu$m and 
1.2\,mm. In a 1-arcmin beam the lensed confusion noise is always predicted to 
be at least 20 per cent of the SZ signal in all three models. Only at a wavelength 
of 2\,mm is the SZ signal expected to be much larger than the confusion noise. 
These results suggest both that the kinematic SZ effect would be difficult to 
detect, and that the thermal SZ effect would be difficult to measure accurately, 
in an observation of a single cluster, unless the population of distant 
dusty star-forming galaxies is very sparse or evolves very weakly. In the light of 
recent observations (SIB) either of these scenarios appears very unlikely. 

\begin{figure}
\begin{center}
\epsfig{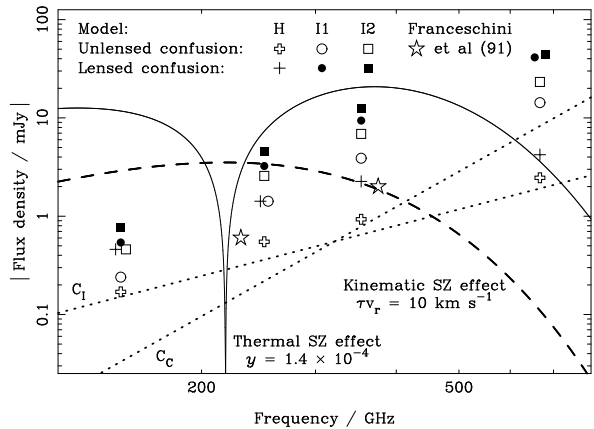}
\end{center}
\caption{The 1-$\sigma$ confusion noise expected in a beam-switched
observation of the SZ effect with and without lensing in a 1-arcmin beam. The 
confusion estimates of Franceschini et al. (1991) and Fischer \& Lange
(1993; dotted lines) are also shown. C$_{\rm C}$ and C$_{\rm I}$ are estimates of 
the confusion due to galactic cirrus emission and {\it IRAS} galaxies respectively. 
C$_{\rm I}$ is very similar to the estimate of confusion noise in Helou \& 
Beichman (1990). 
}
\end{figure}

\begin{figure*}
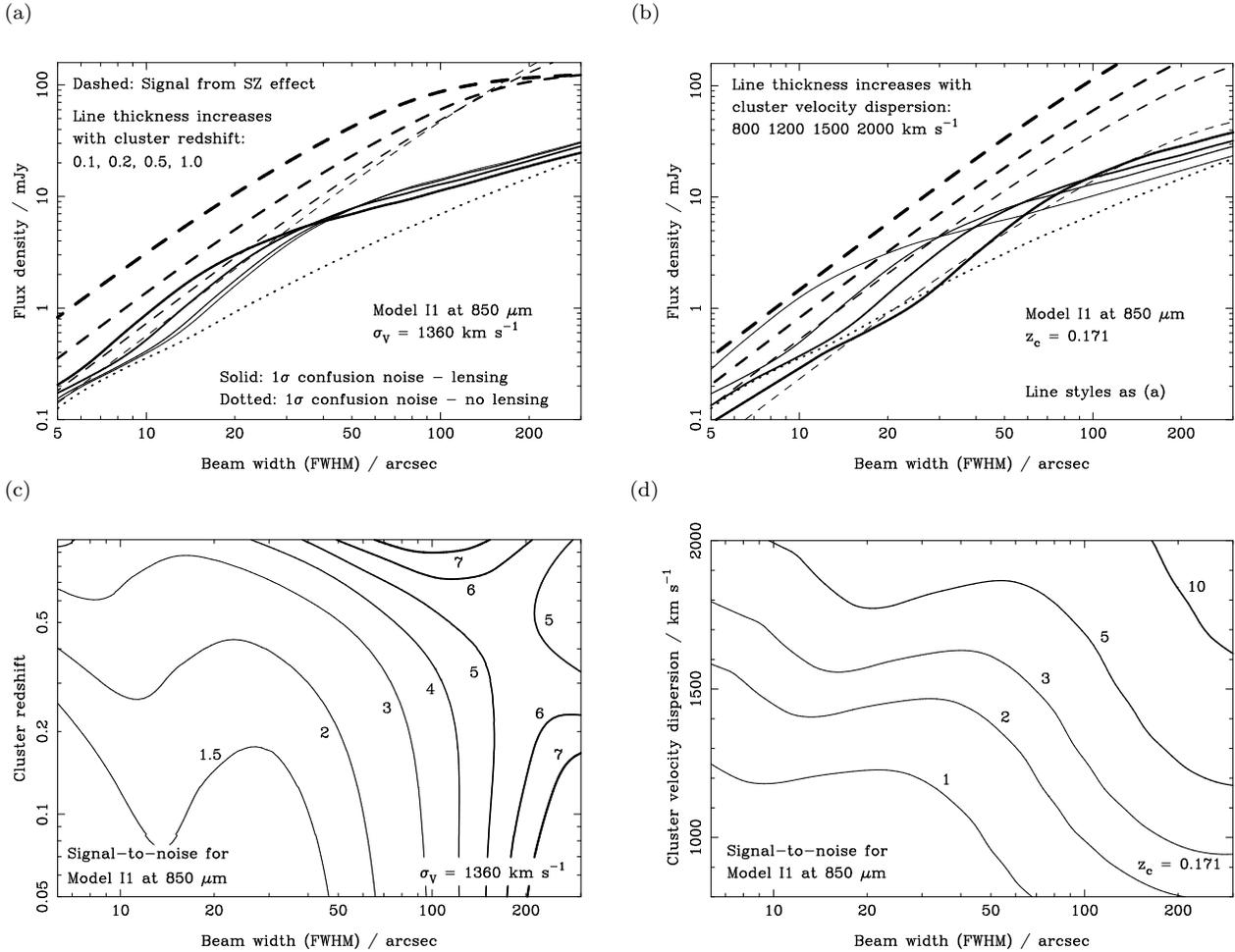

\begin{minipage}{175mm}
(a) \hskip 80mm (b)
\begin{center}
\epsfig{file=Fig5aNov.ps, width=5.05cm,bbllx=325,bblly=30,bburx=762,bbury=514}
\epsfig{file=Fig5bNov.ps, width=5.05cm,bbllx=5,bblly=30,bburx=443,bbury=514}
\end{center}
(c) \hskip 80mm (d)
\begin{center}
\epsfig{file=Fig5cNov.ps, width=5.05cm,bbllx=325,bblly=30,bburx=762,bbury=514}
\epsfig{file=Fig5dNov.ps, width=5.05cm,bbllx=5,bblly=30,bburx=443,bbury=514}
\end{center}
\caption{The effects of the redshift $z_{\rm c}$ and velocity dispersion
$\sigma_{\rm V}$ of a cluster on the SZ signal and source confusion noise 
expected at 850\,$\mu$m in model I1. The properties of the clusters are 
normalised to match the model described in Section~3.2.2, which was
similar to A2218 with $\sigma_{\rm V}$=1360\,km\,s$^{-1}$ and $z_{\rm c}=0.171$. 
The SZ signal and confusion noise as a function of beam-width for a 
cluster at $z_{\rm c}=0.1$, 0.2, 0.5 and 1 are shown in (a); in (b) the same 
quantities are shown for a 
cluster with $\sigma_{\rm V}=800$, 1200, 1500 and 2000\,km\,s$^{-1}$. 
The ratio of the SZ signal to the confusion noise expected for a 
range of values of $z_{\rm c}$ and $\sigma_{\rm V}$ are shown in 
(c) and (d) respectively. 
}
\end{minipage}
\end{figure*} 

\subsection{The mass and redshift of the lensing cluster}

The key parameter that determines the effects of gravitational lensing on the 
confusion noise in an observation is the Einstein radius of the lensing cluster 
$\theta_{\rm E}$, which in turn depends on the redshift $z_{\rm c}$, velocity 
dispersion $\sigma_{\rm V}$ and internal structure of the cluster. If the cluster is 
assumed to be an isothermal sphere, and $D(z,z_{\rm s})$ is the angular diameter 
distance between a redshift $z$ and a source at redshift $z_{\rm s}$, then 
$\theta_{\rm E} \propto D(z_{\rm c},z_{\rm s}) \sigma_{\rm V}^2$ (Paper~1). Note 
that because $z_{\rm s}$ is typically much larger than unity for faint 
submillimetre-wave sources, increasing $z_{\rm c}$ does not significantly reduce 
the surface density of background sources, and so the only significant effect of 
changing $z_{\rm c}$ on the expected confusion due to lensed images is 
introduced through the resulting change in $\theta_{\rm E}$. 
The dependence of the SZ signal from an unresolved cluster on $\sigma_{\rm V}$
and $z_{\rm c}$ was discussed by De Luca, D\'esert \& Puget (1995), using 
models of cluster evolution from Kaiser (1986) and Bartlett \& Silk (1994), and
found to take the form $(1+z_{\rm c}) \sigma_{\rm V}^{10/3} / D(0, z_{\rm c})^2$. 
A beam correction can be added to calculate the SZ signal expected from
clusters that are resolved in smaller observing beams.

The expected dependence of both the confusion noise and the SZ signal at a 
wavelength of 850\,$\mu$m for clusters with $0.05 \le z_{\rm c} \le 1$ and 
800\,km\,s$^{-1} \le \sigma_{\rm v} \le 2000$\,km\,s$^{-1}$ is shown in Figs\,5(a)
\& 5(b) for model I1. The results are normalised to match the predictions for the 
model lensing cluster in the previous sections, which had $z_{\rm c}$ = 0.171 and 
$\sigma_{\rm V}=1360$\,km\,s$^{-1}$. The confusion noise is expected to be 
smaller and larger by a factor of a few in models H and I2 respectively. 
Confusion noise is expected to be less significant as compared with the SZ signal 
in observations of more massive and more distant clusters that are made in 
larger beams, as shown by the signal-to-noise ratios in Figs~5(c) \& 5(d). 

\section{Confusion noise and observations}

\subsection{Existing observations} 

Three ground-based detections of the millimetre-wave SZ effect have been 
published; for Abell~2163 at a wavelength of 2.2\,mm in a 1.4-arcmin-wide beam 
(Wilbanks et~al. 1994), and for Abell~2744 and S1077 at wavelengths of 2 and 
1.2\,mm in 46- and 44-arcsec-wide beams respectively (Andreani et~al. 1996). 
The detected flux densities and estimates of the source confusion noise in 
these observations are listed in Table\,2. 

Silverberg et al. (1997) observed the Coma cluster at four frequencies using
MSAM, a balloon-borne experiment, to search for the submillimetre-wave SZ 
effect. Because of the 28-arcmin beamwidth in this experiment, gravitational 
lensing will have an insignificant effect on the confusion noise in this 
observation. They estimate that CMBR fluctuations at an intensity 
$\Delta T/T \simeq 2\times10^{-5}$ would be a significant source of confusion in 
this observation. Distant dusty galaxies are expected to contribute a significant 
amount of confusion noise in only the highest frequency channel, equivalent 
to $\Delta T/T \simeq 4\times10^{-5}$ at 678\,GHz (BIS).  

A2163 lies at a redshift $z_{\rm c}=0.201$, has a core radius 
$\theta_{\rm c} \simeq 70$\,arcsec (Elbaz, Arnaud \& B\"ohringer 1995), and
$\sigma_{\rm V} = 1680$\,km\,s$^{-1}$ (Markevitch et al. 1996). If the mass
profile of A2163 is assumed to be scaled from that of A2218, then these 
observations indicate an Einstein radius $\theta_{\rm E} \simeq 24$\,arcsec, 
which leads to the estimates of confusion noise listed in Table\,2. 
Hence, although we expect lensing to increase the level of source confusion 
noise in this observation, it should not exceed about 1 per cent of the detected 
SZ signal in this cluster. Upgraded versions of the SuZIE instrument
used in this observation (Holzapfel et al. 1997) continue to offer excellent 
performance for detecting the millimetre/submillimetre-wave SZ effect. 

The SZ signals listed in Table~2 for A2744 and S1077, at redshifts 
$z_{\rm c} = 0.308$ and 0.312 respectively, are derived from the $y$-parameters 
quoted by Andreani et al. (1996), and depend only weakly on the assumed core 
radii of these clusters. The differential signal from this beam-switched
observation would be smaller by about 30 to 40 per cent due to the signal
from the SZ effect that remains in the reference beams 135\,arcsec 
away on the sky. Estimates of the Einstein radii of A2744 and S1077 are required 
in order to determine the confusion noise expected in these observations. The 
temperatures of 7\,keV quoted for the intracluster gas in both A2744 and S1077 
match that of A2218 (Squires et al. 1996). Hence, if self-similar cluster evolution 
is valid (Kaiser 1986), then $\theta_{\rm E} \simeq 13$\,arcsec for both clusters. 
However, the temperature of 12\,keV has recently been determined for A2744 
(Mushotzky \& Scharf 1997), in which case $\theta_{\rm E} \simeq 26$\,arcsec. The 
values of confusion noise that are predicted assuming values of 
$\theta_{\rm E} \simeq 13$ and 26\,arcsec for S1077 and A2744 respectively are 
listed in Table\,2. 
These values are greater than about 1 and 10 per cent of the SZ signal at 
wavelengths of 2 and 1.2\,mm respectively in all three models of galaxy
evolution. In models I1 and I2 the unlensed source confusion at a wavelength of 
1.2\,mm is expected to be several times larger than the value of 0.4\,mJy 
suggested by Andreani et al. (1996). A further increase in confusion noise due to 
bright lensed images in the reference beam is very unlikely, because the 
reference beams are separated by 135\,arcsec from the cluster core.

We expect the confusion noise in existing observations of the millimetre-wave SZ 
effect to be about 1 and 10 per cent of the SZ signal at wavelengths of about 
2\,mm and 1.2\,mm respectively, and so these detections are unlikely to be 
artifacts of confusion noise. If a large 
number of detections of the SZ effect could be obtained at 1.2\,mm then a careful 
analysis of the sources of noise could be used to estimate the amount of source 
confusion present, and so limit the form of evolution of distant dusty galaxies. 
Observations at several wavelengths would assist this analysis, as the amplitude 
of the confusing signal at each wavelength should be correlated for each cluster. 

\subsection{Future observations}

A range of new instruments and telescopes could detect the 
millimetre/submillimetre-wave SZ effect. Suitable space-borne instruments 
include ESA's {\it Far InfraRed and Submillimetre Telescope (FIRST)}, a 
3.5-m submillimetre-wave/far-infrared telescope (Pilbratt 1997), and 
{\it Planck Surveyor} (formerly {\it COBRAS/SAMBA}; Bersanelli et al. 1996), a 
CMBR imaging satellite with a 1.5\,m$\times1.292$\,m aperture. 
Suitable ground-based instruments include the SCUBA camera at the 
James Clerk Maxwell Telescope (Cunningham et al. 1994), the US-Mexican 50-m 
Large Millimetre Telescope (LMT) (Cortes-Medellin \& Goldsmith 1994) and 
large millimetre interferometer arrays (MIAs; Downes 1994). Observations by 
balloon-borne experiments, such as MSAM (Silverberg 1997) and BOOMERANG 
(Lange et al. 1995), are also capable of detecting the submillimetre-wave SZ
effect. However, these instruments have beamwidths of order several tens of 
arcminutes and so the effects of gravitational lensing are unimportant. Applicable
estimates of unlensed source confusion for these experiments can be found in BIS. 

\subsubsection{\it FIRST} 

At present, the longest observing wavelength of {\it FIRST}'s bolometer 
instrument is 500\,$\mu$m (Griffin 1997), and so the instrument's spectral range is
not optimized to detect the peak of the signal due to the SZ effect at a
wavelength of about 800\,$\mu$m (Fig.\,1). {\it FIRST}'s angular resolution 
is finer than about 40\,arcsec at 500\,$\mu$m, and so any observations of the SZ 
effect are likely to be affected by source confusion. The expected flux density 
from the core of a cluster with $y=10^{-4}$ in a 40-arcsec beam is about 4\,mJy, 
as compared with estimated lensed and unlensed 1$\sigma$ source confusion
noise values of at least 6 and 2\,mJy\,beam$^{-1}$ in model I1, Fig.\,5(a). 
{\it FIRST} should reach a 1$\sigma$ sensitivity limit of about 
1\,mJy\,beam$^{-1}$ at 500\,$\mu$m in a 25-arcmin$^2$ field in a 6-hour 
integration (Griffin 1997), and so a detection would be possible in a reasonable 
integration time in the absence of source confusion. However, without 
resolving and subtracting the flux density from confusing sources obtaining a
reliable detection for an individual cluster using {\it FIRST} would be difficult.

\subsubsection{\it Planck Surveyor} 

In a 1-year all-sky survey {\it Planck Surveyor} should detect the SZ effect in 
about $1.5\times10^4$ clusters with $y > 8\times10^{-5}$ and core radii 
subtending about 1\,arcmin (Bersanelli et al. 1996). The kinematic SZ signal of the 
brighter clusters in the sample could be used to determine their peculiar 
velocities to an accuracy of about 100 to 300\,km\,s$^{-1}$ (Haehnelt \& 
Tegmark 1996). Similar predictions have recently been made by Aghanim et al.
(1997). A 1.5\,m$\times1.292$\,m aperture corresponds to a beam-width 
of about 3\,arcmin at 850\,$\mu$m, and so gravitationally lensed source 
confusion is not expected to degrade the performance of {\it Planck Surveyor} 
by a large amount (Fig.\,5). The unlensed confusion noise expected for 
{\it Planck Surveyor} is discussed by BIS: in the 353-GHz channel, in which
the thermal SZ signal is expected to be largest, a 1$\sigma$ noise level of 
about 12\,mJy is expected in each 19-arcmin$^2$ pixel. The nominal 1$\sigma$
sensitivity of the {\it Planck} survey is comparable, at 16\,mJy\,pixel$^{-1}$
(Bersanelli et al. 1996).

\begin{table}
\caption{Estimates of source confusion noise in existing observations of the SZ
effect (Wilbanks et~al. 1994; Andreani et al. 1996). The results are derived in
Section~4.1. The observing wavelength $\lambda$ and reported SZ signal 
$S_{\rm SZ}$ are also listed.}
\begin{tabular}{ p{0.85cm}p{0.55cm}p{1.3cm}p{4.5cm} } 
Cluster & $\lambda$ / & $S_{\rm SZ}$ / & 1-$\sigma$ confusion noise: \\
            & mm & mJy & lensed (unlensed) / mJy \\
\end{tabular}
\begin{tabular}{p{0.85cm}p{0.55cm}p{0.45cm}p{1.4cm}p{1.4cm}p{1.4cm} } 
 & & & Model I1 & Model I2 & Model H \\
\noalign{\vskip 4pt}
A2163 & 2.2 & -68 & 0.75 (0.36) & 1.04 (0.60) & 0.65 (0.25) \\
\noalign{\vskip 2pt}
A2744 & 2.0 & -19 & 0.33 (0.18) & 0.41 (0.32) & 0.34 (0.13) \\
           & 1.2 & 13 & 1.80 (0.96) & 2.27 (1.75) & 1.00 (0.38) \\
S1077 & 2.0 & -26 & 0.41 (0.18) & 0.58 (0.32) & 0.36 (0.13) \\
           & 1.2 & 18 & 2.34 (0.96) & 3.16 (1.75) & 1.06 (0.38) \\
\end{tabular}
\end{table}

\subsubsection{SCUBA} 
 
In a 30-hour integration in a 10-arcmin$^2$ field SCUBA should be able to 
detect point sources with flux densities of about 7 and 40\,mJy at a signal to 
noise ratio of 3$\sigma$, given the currently attainable sensitivities (SIB, JCMT
web page). These 
sensitivities match the thermal SZ signal from the core of clusters with 
$y$-parameters of $8.4\times10^{-4}$ and 0.12 at 850 and 450\,$\mu$m 
respectively. SCUBA could hence detect a cluster with $y=1.4\times10^{-4}$ at 
a signal to noise ratio of about 0.5$\sigma$\,beam$^{-1}$ at the centre of the 
field at 850\,$\mu$m; however, the cluster could not be detected at 450\,$\mu$m. 
The resulting 850-$\mu$m map with an angular resolution of about 13\,arcsec 
would contain about 200 resolution elements. The surface brightness of the 
SZ signal would still have 40 per cent of its central value at the edge of the field 
if $\theta_{\rm c} \simeq 50$\,arcsec, and so the signal to noise ratio for the 
thermal SZ effect in the observation would be about 3$\sigma$. The 
corresponding kinematic effect would not be detectable, reaching a 
signal-to-noise ratio of only 0.5 if $\tau v_{\rm r}=10$\,km\,s$^{-1}$.
In general, an observation of a 10-arcmin$^2$ field lasting $t$\,hours should 
detect the thermal SZ effect from a cluster with $\theta_{\rm c}\simeq50$\,arcsec
at a signal to noise ratio of about $0.4[y/10^{-4}]t^{1/2}$. In a fixed integration 
time, the signal to noise ratio for an observation of the SZ effect in a circular 
field centred on the cluster is maximized when the radius of the field is twice 
that of the cluster core. 

SCUBA will allow limited source subtraction of the brightest lensed images in 
the field, as discussed in Paper\,1 and recently confirmed by SIB. However, 
confusion noise from fainter images could still be significant. In 
order to subtract a larger fraction of lensed images and reduce the confusion
noise, observations using either the LMT, with an angular resolution limit of
about  5\,arcsec at a wavelength of 1\,mm, or a large MIA, with sub-arcsecond 
resolution, would be required.

\subsubsection{Millimetre interferometer arrays} 

The performance of large MIAs (Downes 1994) was discussed in Paper~1 in the 
context of detecting lensed images in the fields of clusters. The large collecting 
areas and very sensitive receivers of these instruments will allow very faint
images to be detected at sub-arcsecond angular resolution. However, the 
instantaneous field-of-view of a large interferometer is limited to the primary 
beam area of its individual antennas, which is about 0.4\,arcmin$^2$ for an array 
of 8-m antennas at a wavelength of 1.3\,mm, and so observations of fields that 
are several arcminutes in extent would require the accurate mosaicing 
(Cornwell 1992) of many different fields. A mosaic of about 25 primary beam
areas would be required in order to make the most sensitive detection of the SZ 
effect at 850\,$\mu$m from a cluster with $\theta_{\rm c} \simeq 50$\,arcsec in a 
9-arcmin$^2$ field. Fewer mosaiced fields would be required at longer 
wavelengths, for which the primary beam area is larger. At a wavelength of 3\,mm 
the primary beam area of an array of 8-m antenna would be about 2\,arcmin$^2$, 
and so a mosaic of only 5 beams would be required to map the cluster. Hence, 
the most efficient strategy to detect the SZ effect using an MIA would probably 
be to observe in the millimetre rather than the submillimetre waveband.

In a compact array configuration, the proposed Large Southern Array (Downes 
1996) should be able to map a 9-arcmin$^2$ field down to a 3-$\sigma$ 
sensitivity of about 0.13\,mJy in each 10-arcsec$^2$ synthesised beam at a 
wavelength of 1.3\,mm in a 4-hour integration. The Millimeter Array (Brown 
1996) and Large Millimetre and Submillimetre Array (Ishiguro et al. 1992) should 
offer a similar performance. An MIA could hence detect the brightest lensed 
images in a rich cluster in a matter of hours, and would be ideal for carrying out 
sensitive source subtraction in the millimetre/submillimetre waveband. If an MIA is
assumed to have a constant point source sensitivity, then the flux density limit
at any wavelength can be estimated by scaling the value given above at
1.3\,mm by the areas of the primary and synthesized beams at other wavelengths. 
In this case, the thermal SZ effect from a cluster with $y=1.4\times10^{-4}$ and 
$\theta_{\rm c}=50$\,arcsec could be detected by an MIA with signal-to-noise 
ratios of about 40, 15, 2 and 6$\sigma$ in 1-hour integrations at 
wavelengths of 3, 2, 1.3 and 0.85\,mm respectively. An MIA could hence obtain a
very significant detection of the millimetre-wave SZ effect. Confusing 
sources could be subtracted accurately using observations on longer baselines, 
and so a very detailed image of the millimetre-wave SZ effect could be produced. 

\section{Conclusions}

The effects of the lensed images of distant dusty star-forming 
galaxies on observations of the SZ effect in the millimetre/submillimetre 
waveband, in which distant sources are intrinsically bright have been 
discussed. Such images have
recently been detected by Smail et al. (1997). 
\begin{enumerate} 
\item 
Gravitational lensing of distant galaxies is expected to increase source confusion 
noise in millimetre/submillimetre-wave observations of the SZ effect. The most 
pronounced increase, by a factor of about 3, is expected in observing beams that 
are matched to the Einstein radius of the lensing cluster. This increased estimate 
of confusion noise is not sufficiently large to affect the interpretation of existing 
millimetre-wave observations of the SZ effect; however, it could provide a  
significant source of noise in future observations on arcminute scales, and 
the brightest sources may need to be subtracted in order to measure the SZ 
effect accurately. 
\item 
The effect of gravitational lensing on confusion noise in the fields of clusters 
depends on both the form of the population and evolution of distant star-forming 
galaxies, and the surface density of the lensing cluster. In general, a larger 
signal-to-noise ratio would be obtained from an observation of a more massive 
cluster, at a larger redshift, in an observing beam that is several times larger 
than the Einstein radius of the cluster. 
\item 
Future millimetre/submillimetre-wave observations of the SZ effect using
ground-based and space-borne instruments will be affected by confusion in
different ways. The angular resolution of {\it Planck Surveyor} is probably 
sufficiently coarse to avoid significantly increased source confusion due to 
lensing in clusters. Observations using telescopes with smaller beams, such as 
{\it FIRST} and ground-based single-antenna telescopes, will be more seriously 
affected. 
Large millimetre interferometer arrays will be very valuable instruments for 
carrying out source subtraction and will be able to detect the SZ effect directly, 
most easily in the millimetre waveband. 

\end{enumerate} 

\section*{Acknowledgements}
I would like to thank Alastair Edge for his prompt refereeing of this paper, and 
Malcolm Longair for useful discussions during its preparation.

\end{document}